\def\GranadaIns{Instituto Carlos I de F\'\i sica Te\'orica y
Computacional,
Facultad de Ciencias, \\ Universidad de Granada, Campus de
Fuentenueva,
Granada, Spain E-18002}

\def\Valencia{Instituto de F\'\i sica Corpuscular, Centro Mixto Universidad de 
Valencia-CSIC, Burjasot, Valencia, Spain E-46100}

\def\Napoli{Dipartimento di Scienze Fisiche, Universit\`a degli studi di 
Napoli,\\ Mostra d'Oltremare, Pad. 19, 80125 Napoli, Italy}

\def\nn{\nonumber}
\def\cinf{$c\rightarrow\infty\;$}

\def\w {\omega}
\def\ni{\noindent}

\def\Gt{$\widetilde{G}$}
\def\Nl{N_\lambda}
\def\CNln{C^{(N,\lambda)}_{n}}
\def\CNlm{C^{(N,\lambda)}_{m}}
\def\CpNln{C'^{(N,\lambda)}_{n}}
\def\ii{{\rm i}}
\def\e{{\rm e}}
\def\d{{\rm d}}
\def\HNln{H^{(N,\lambda)}_{n}}
\def\medio{\frac{1}{2}}

\documentstyle[12pt]{article}

\begin{document}
 



\begin{center} 
{\Large {\bf A Perturbative Approach to the \\ Relativistic 
             Harmonic Oscillator}}
\end{center}

\bigskip
\bigskip

\centerline{ J. Guerrero$^{1,2}$ and V. Aldaya$^{1,3}$}

\bigskip
\centerline{January 29, 1998}
\bigskip

\footnotetext[1]{\GranadaIns} 
\footnotetext[2]{\Napoli}  
\footnotetext[3]{\Valencia}


\bigskip

\begin{center}
{\bf Abstract}
\end{center}

\small

\begin{list}{}{\setlength{\leftmargin}{3pc}\setlength{\rightmargin}{3pc}}
\item A quantum realization of the Relativistic Harmonic Oscillator is  
realized in
terms of the spatial variable $x$ and ${\d\over \d x}$ (the minimal
canonical representation). The eigenstates of the Hamiltonian operator 
are found (at lower order) by using a perturbation expansion in the 
constant $c^{-1}$. Unlike the Foldy-Wouthuysen transformed version of
the relativistic hydrogen atom, conventional perturbation theory cannot be 
applied and a perturbation of the scalar product itself is required.
\end{list}

\normalsize

\vskip 1cm

\section{Introduction}
The Relativistic Harmonic Oscillator is probably the simplest
relativistic system containing bound states, yet it exhibits the typical
problems
of Relativistic Quantum Mechanics. Many papers have been devoted to
the
solution of this relativistc system 
\cite{Yukawa,Mir-Kasimov,Feynman,Kim78,Arshansky} 
although the first question is probably to define what we understand by
a Relativistic Harmonic Oscillator.

In a previous paper \cite{Aldaya91}, we adopted an algebraic method for
both defining and
solving such an oscillator equation which started with the Lie operator algebra
\begin{equation}
[\hat{E},\hat{x}]=-\ii\frac{\hbar}{m}\hat{p},\,\,\,\,\,\,
[\hat{E},\hat{p}]=\ii m\omega^{2}\hbar\hat{x},\,\,\,\,\,\,
[\hat{x},\hat{p}]=\ii\hbar(1+\frac{1}{mc^{2}}\hat{E})\, ,
\label{Epxalg}
\end{equation}
\noindent which is an affine version of the Lie algebra of the 1+1 anti-de 
Sitter group
($\approx sl(2,R)$) and reproduces the appropriate limits (in the sense of 
In\"on\"u and Wigner group contractions \cite{Inonu-Wigner}) as
$\omega\rightarrow 0$ (going to the Poincar\'e group in 1+1D) and/or \cinf 
(going to the Harmonic Oscillator group), although in this paper we will be 
concerned only with 
the \cinf limit, i.e. the one leading to the Harmonic Oscillator. The 
solution was given through a
{\it manifestly covariant} representation of the corresponding
group (i.e. the wave functions are solutions of a Klein-Gordon-like equation).
The energy eigenfunctions in configuration space consist of a
general
weight function (the vacuum) which converges to a Gaussian in the limit \cinf ,
a specific power of the vacuum (reflecting the explicit
dependence
on time of the manifestly covariant representation) which reduces to unity
in this
limit, the usual time-dependent phase factor $\exp(-\ii E_nt)$ and a
polynomial leading
to the corresponding non-relativistic Hermite polynomial as \cinf. As
a
consequence of the explicit dependence on time, the scalar product
was defined
by using the invariant measure $\d x\d t$. These representations correspond
to a realization in the real variables $(x,t)$ of the Discrete series 
$D^+(k)$ of the $SL(2,R)$
group \cite{Bargmann,Lang} for Bargmann index $k=N\equiv 
\frac{mc^2}{\hbar\omega}$.

In this paper we study the problem appearing in the
configuration-space
representation when the time dependence is factorized out in the
search for
a {\it minimal, canonical realization} (given in terms of only $x$ and
$\frac{\d}{\d x}$) in a
way similar to that of the relativistic hydrogen atom. In this last case,
a
Foldy-Wouthuysen transformation leads to a Hamiltonian containing
higher-order
relativistic corrections to the non-relativistic
Hamiltonian \cite{Landau}.
Ordinary perturbation theory provides the energy levels as a
power series
in $c^{-1}$ of the exact values already obtained from the
manifestly covariant (Dirac) equation.

In the present case, however, the situation is quite different because the 
terms in the Hamiltonian that are considered here as higher-order relativistic
corrections
to the non-relativistic Hamiltonian are not Hermitian with
respect to the 
non-relativistic scalar product (with measure $\d x$) and a
perturbation
of the scalar product itself is required. These higher-order relativistic 
corrections should be understood as a power expansion in $1/k$ of the
representations of the Discrete series for large Bargmann indices 
$k=N=\frac{mc^2}{\hbar\omega}$.

Our group quantization method \cite{Aldaya82} essentially consists in 
exponentiating the abstract algebra (either Poisson or operator algebra) 
of basic quantities defining a given physical system, usually a 
$u(1)$-centrally extended algebra $\tilde{\cal G}$, and considering in the
resulting group (\Gt) the two mutually commuting sets of vectors fields, 
$\chi^R$(\Gt) and $\chi^L$(\Gt) which generate the left and right action of 
the group on itself. One set of vector fields will constitute the operators of
the theory, while the other can be used to fully reduce the action of the
former. To be a bit more precise, we consider the subspace of complex 
functions on \Gt, ${\cal F}$(\Gt),
that satisfy the
$U(1)$-equivariance condition, $\Xi\Psi=\ii\Psi$, where $\Xi$ is the
central generator $\tilde{X}^{L}_{\zeta}=\tilde{X}^{R}_{\zeta}$ and $\zeta\in
U(1)$.
The right-invariant vector fields (generating the finite left action)
$\tilde{X}^R\in\chi^R$(\Gt) act on ${\cal F}$(\Gt) as a reducible representation
(corresponding to the Bohr-Sommerfeld quantization). The reduction defining 
the (true)
quantization is achieved by imposing on ${\cal F}$(\Gt) the so-called
polarization condition in terms of a subalgebra ${\cal P}$  of
left-invariant
vector fields, $\tilde{X}^L\Psi=0, \forall\tilde{X}^L\in{\cal P}$,
which preserve the action of $\tilde{X}^R$ $([\tilde{X}^R,\tilde{X}^L]=0)$.
The polarization ${\cal P}$,
originally defined as a maximal left subalgebra of \Gt\  containing the
kernel of the Lie algebra cocycle $\Sigma$ and excluding the central generator
$\Xi$, \cite{Aldaya82}
can be further generalized by allowing it to contain
operators in the
left enveloping algebra $U\chi^L$(\Gt) \cite{Loll,Marmo}.

The paper is organized as follows: In section 2 the exact solutions for the
Relativistic Harmonic Oscillator, firstly given in \cite{Aldaya91} and 
obtained later through a second order polarization
in \cite{Oscilata}, are given. A reduction to a minimal representation (in 
terms of $x$ and ${d \over dx}$) is tempted through a naive elimination of the
time variable and the substitution $i\hbar{\partial \over \partial t} 
\rightarrow E_n -mc^2$, but this leads to a theory which is not unitary. The 
solution is achieved by modifying the scalar product and the operators in a 
consistent though not quite well understood way.
In section 3 a perturbative approach to the problem of the reduction is 
proposed. From an exact infinite-order polarization (which represents the
``square root" of the Casimir leading to the Klein-Gordon equation) we obtain
a perturbative expansion for a Schr\"odinger-like equation, where the 
zeroth-order Hamiltonian is the non-relativistic Harmonic Oscillator 
Hamiltonian.
From this Schr\"odinger equation we inmediately realize that the minimal 
representation is not unitary because the perturbed Hamiltonian (as opposite
to the case of the hydrogen atom) is not Hermitian with respect to the measure
$dx$ of the non-perturbed theory. Thus a perturbation of the scalar product 
itself is proposed, and the solution coincides (at lower order in powers of
$1\over c^2$) with the exact solution proposed in section 2.

\section{The Relativistic Harmonic Oscillator}

To quantize the physical system characterized by the algebra
(\ref{Epxalg}),
we must exponentiate this algebra and derive left and right vector
fields. The left ones are\footnote{They were called in Ref. \cite{Aldaya91}
$L_t, L_p, L_x$, respectively} (see \cite{Oscilata} for the expression on the
right-invariant vector fields):
\begin{eqnarray}
 \tilde{X}^{L}_{t}&=&\frac{p}{m}\frac{\partial}{\partial
x}-m\omega^{2}x
      \frac{\partial}{\partial p}+\frac{P_{0}}{mc\alpha^{2}}
   \frac{\partial}{\partial t} \nn \\
 \tilde{X}^{L}_{p}&=&\frac{P_{0}}{mc}\frac{\partial}{\partial p}+
      \frac{mcx}{P_{0}+mc}\frac{1}{\hbar}\Xi \nn \\
 \tilde{X}^{L}_{x}&=&\frac{P_{0}}{mc}\frac{\partial}{\partial x}+
       \frac{p}{mc^{2}\alpha^{2}}\frac{\partial}{\partial t}-
       \frac{pmc}{P_{0}+mc}\frac{1}{\hbar}\Xi  \, ,
 \end{eqnarray}
\noindent where
$P_{0}\equiv\sqrt{mc^{2}+p^{2}+m^{2}\omega^{2}x^{2}}$ and
$\alpha\equiv\sqrt{1+\omega^2c^{-2}x^2}$. 

As was mentioned in the introduction, the representation given by the 
right-invariant vector fields acting on complex wave functions on the group
\Gt\ is reducible. Thus, we need to impose conditions on the wave 
functions (polarization conditions) in order to reduce the representation
space, and this is achieved by a polarization subalgebra of 
left-invariant vector fields. There is a first-order polarization, leading to
the Bargmann-Fock representation \cite{Oscilata}, but in this paper we are
interested in the configuration-space representation, and for this purpose
we need a higher-order polarization \cite{Loll,Marmo}.

There is a second-order polarization which leads to the manifestly
covariant representation:
\begin{equation}
<\tilde{X}^{HO}_{t}\equiv(\tilde{X}^{L}_{t})^{2}-
c^{2}(\tilde{X}^{L}_{x})^{2}
  +\frac{2\ii mc^{2}}{\hbar}
    \tilde{X}^{L}_{t}+\lambda\frac{\ii
mc^{2}\omega}{\hbar}\Xi,\;\tilde{X}^{L}_{p}>\,,\label{GHOP}\end{equation}
\noindent where the numerical parameter $\lambda$ is arbitrary but can
be chosen
to yield the results previously obtained \cite{Aldaya91}. Imposing
the $U(1)$-equivariance condition and solving
the equation $\tilde{X}^{L}_{p}\Psi=0$ allows us to factor out the common 
$\zeta$ and $p$-dependence. The remaining equation is a Klein-Gordon-like 
equation for $\varphi(x,t)$:
\begin{equation}
\frac{1}{\alpha^{2}}
\frac{\partial^{2}\varphi}{\partial t^{2}}-
     \frac{2imc^{2}}{\hbar\alpha^{2}}\frac{\partial\varphi}{\partial t}-
     2\omega^{2}x\frac{\partial\varphi}{\partial x}-
     c^{2}\alpha^{2}\frac{\partial^{2}\varphi}{\partial x^{2}}- 
    \frac{m^{2}c^{4}}{\hbar^{2}\alpha^{2}}\varphi- 
    \lambda\frac{mc^{2}\omega}{\hbar}\varphi + \frac{m^2c^4}{\hbar^2}
\varphi=0\,.
\label{Klein-Gordon}
\end{equation}

 \ni In this equation the wave funtion $\varphi(x,t)$ has the rest mass 
substracted. If we restore it, defining 
$\tilde{\varphi}=e^{\frac{i}{\hbar}mc^2t}\varphi$, we obtain the more 
standard expression:
\begin{equation}
\left(\Box + \frac{m^2c^2}{\hbar^2} + \chi R\right)\tilde{\varphi} = 0\,,
\end{equation}

\ni where $\Box\equiv\frac{1}{c^{2}\alpha^{2}}\frac{\partial^{2}}
{\partial t^{2}}- \frac{2\omega^{2}x}{c^{2}}\frac{\partial}{\partial x}-
\alpha^{2}\frac{\partial^{2}}{\partial x^{2}}$ is the D`Alembert operator on 
anti-de Sitter space-time (in 1+1D), $R\equiv -2\frac{\w^2}{c^2}$ is the 
scalar curvature and $\chi\equiv \frac{N\lambda}{2}$ is a parameter 
providing the coupling of the scalar field to the gravitational field
(see \cite{Birrell}). 

The normalized (positive-energy) solutions to the Klein-Gordon-like 
equation are:
\begin{equation}
\Psi^{(N,\lambda)}_{n}(x,t)\equiv \CNln \e^{-\ii b_{n}\omega
t}\alpha^{-c_{n}}\HNln(x)\,, \label{exactas}
\end{equation}
\ni where $b_{n}=c_{n}=c_{0}+n\equiv\frac{1}{2}+N_{\lambda}+n \equiv
  E_n^{(N,\lambda)}/\hbar\w$, $N_{\lambda}\equiv\frac{1}{2}
\sqrt{1+4N(N-\lambda)}$,
\begin{equation}
\CNln=\sqrt{\frac{\w}{2\pi}}
 \left(\frac{m\w }{\hbar\pi}\right)^{\frac{1}{4}}
 \frac{1}{2^{n/2}\sqrt{n!}}
             \sqrt{\frac{\Gamma(2\Nl+1)(2N)^n}{\Gamma(2\Nl+n+1)}}
 \sqrt{\frac{\Gamma(\Nl+\frac{1}{2})}{\sqrt{N}\Gamma(\Nl)}} \, ,
 \end{equation}
\ni $N=\frac{mc^2}{\hbar\omega}$,
and the {\it Relativistic Hermite polynomials} \cite{Aldaya91,Oscilata},
$\HNln$, satisfy:
 \begin{equation}
 (1+\frac{\xi^{2}}{N})\frac{\d^{2}}{\d\xi^{2}}\HNln-\frac{2}{N}
      (N_{\lambda}+n-\frac{1}{2})\xi\frac{\d}{\d\xi}\HNln+
      \frac{n}{N}(2N_{\lambda}+n)\HNln=0 \, ,
 \label{HEq}
 \end{equation}
\ni where $\xi\equiv\sqrt{\frac{m\omega}{\hbar}}x$ (see \cite{Nagel} for the
relation between the Relativistic Hermite Polynomials and the Gegenbauer
polynomials).

The wave functions are 
orthonormal according to the $t$-$x$ scalar
product
\begin{eqnarray}
<\Psi^{(N,\lambda)}_{n}|\Psi^{(N,\lambda)}_{m}> &=& \CNln\CNlm\int \d
x \d t \e^{-\ii(m-n)\w t}
\alpha^{-(1+2\Nl+n+m)} \times \nn \\
 & & \HNln H^{(N,\lambda)}_{m}
= \delta_{nm} \, , \label{esc1}
\end{eqnarray}
\noindent the measure of which comes from the invariant volume 
$P_0^{-1}\d p\d x\d t$ after a simple but non-trivial 
regularization of the p-integration \cite{Oscilata}.

The annihilation and creation operators (see \cite{Oscilata}), when acting
on $\varphi(x,t)$, have the form:
\begin{eqnarray}
\hat{Z} &=& \sqrt{\frac{\hbar}{2m\w}} e^{i\w t}\left[ 
\alpha \frac{\partial}{\partial x} + 
i\frac{\w x}{c^2\alpha}\frac{\partial}{\partial t} + 
\frac{m\w x}{\hbar\alpha}\right] \nn\\
\hat{Z}^{\dag} &=& \sqrt{\frac{\hbar}{2m\w}} e^{-i\w t}\left[ 
-\alpha \frac{\partial}{\partial x} +
i\frac{\w x}{c^2\alpha}\frac{\partial}{\partial t} + 
\frac{m\w x}{\hbar\alpha}\right]\,. 
\end{eqnarray}

These operators are the adjoint of each other with respect to the scalar 
product (\ref{esc1}). Their action on (\ref{exactas}) is:
\begin{eqnarray}
\hat{Z} \Psi^{(N,\lambda)}_n &=& \sqrt{ n\,\frac{2\Nl+n}{2N}}\,
                                 \Psi^{(N,\lambda)}_{n-1} \nn\\
\hat{Z}^{\dag}\Psi^{(N,\lambda)}_n &=& \sqrt{(n+1)\,\frac{2\Nl+n+1}{2N}}\,
                                 \Psi^{(N,\lambda)}_{n+1} \,.
\end{eqnarray}

The representations here obtained belong to the Discrete series $D^+(k)$ of
$sl(2,R)$ with Bargmann index $k=N=\frac{mc^2}{\hbar\omega}$. Only for
half-integer values of $k>\frac{1}{2}$ they exponentiate to a univalued 
representation of the group $SL(2,R)$ (the rest of the values of $k$ provide,
however, univalued representations of the universal covering group of 
$SL(2,R)$).

A problem arises, however, when one tries to factorize out the time
dependence to obtain a {\it minimal} representation. The functions
$\alpha^{-c_{n}}H^{(N,{\lambda})}_{n}$ are no longer orthogonal unless
we modify the scalar product in the form
$\int \d x \rightarrow \int\d x\alpha^{-2}$ (for a discussion on the 
non-fully understood modified scalar product see \cite{Oscilata}).

With this new scalar product, the normalized wave functions are:
\begin{equation}
\Psi'^{(N,\lambda)}_{n}(x)\equiv \CpNln \alpha^{-c_{n}}
\HNln(x)\,, \label{exactas2}
\end{equation}

\ni with 
\begin{equation}
\CpNln=\sqrt{\frac{\w}{2\pi}}
 \left(\frac{m\w }{\hbar\pi}\right)^{\frac{1}{4}}
 \frac{1}{2^{n/2}\sqrt{n!}}
             \sqrt{\frac{\Gamma(2\Nl+1)(2N)^n}{\Gamma(2\Nl+n+1)}}
 \sqrt{\frac{\Nl+n+\medio}{\Nl+\medio}}
 \sqrt{\frac{\Gamma(\Nl+\frac{3}{2})}{\sqrt{N}\Gamma(\Nl+1)}} \, ,
 \end{equation}

Neither the operators $\hat{Z}$ and $\hat{Z}^{\dag}$ are adjoint to each other
with respect to the new scalar product, so they must be appropiately 
corrected. The relation between the corrected operators and the old ones, 
when acting on the energy eigenfunctions, is given by:
\begin{eqnarray}
\hat{Z}' &=& e^{-i\w t} \sqrt{\frac{\Nl+n-\medio}{\Nl+n+\medio}}
                  \,\, \hat{Z} \nn\\
\hat{Z}'^{\dag} &=& e^{i\w t} \sqrt{\frac{\Nl+n+\frac{3}{2}}{\Nl+n+\medio}}
                  \,\, \hat{Z}^{\dag} \,.
\end{eqnarray}

\ni These expresions are the generalization to arbitrary $\lambda$ of the
ones given in \cite{Oscilata} for $\lambda=1$. 

These results suggest that there is a unitary transformation $\hat{U}$
relating the manifestly covariant and the minimal representatations, i.e.
$\Psi'^{(N,\lambda)}_n(x) = \hat{U} \Psi^{(N,\lambda)}(x,t)$. This 
transformation is of the form:
\begin{equation}
\hat{U} = e^{\frac{i}{\hbar}t(E_n^{(N,\lambda)}-mc^2)}
           \sqrt{\frac{E_n^{(N,\lambda)}}{\hbar\w \Nl}}\,,\label{U}
\end{equation}

\ni when acting on energy eigenfunctions. Having into account that the 
Hamiltonian $\hat{H}\equiv i\hbar {\partial \over \partial t}$ satisfies
$\hat{H}\,\Psi^{(N,\lambda)}_n(x,t) = E_n^{(N,\lambda)}\,
\Psi^{(N,\lambda)}_n(x,t)$, we can 
write $\hat{U}$ as:
\begin{equation}
\hat{U} = e^{\frac{i}{\hbar}t(\hat{H}-mc^2)}
           \sqrt{\frac{\hat{H}}{\hbar\w \Nl}}\,,\label{U2}
\end{equation}

\ni when acting on an arbitrary function. With this expresion we can obtain the
form of the operators $\hat{Z}'$ and $\hat{Z}'^{\dag}$ when acting on arbitrary
functions, not only energy eigenfunctions, simply 
transforming them by $\hat{U}$: $\hat{Z}'= \hat{U}\hat{Z}\hat{U}^{-1}\,,\,
\hat{Z}'^{\dag}= \hat{U}\hat{Z}^{\dag}\hat{U}^{-1}$.
 
One of the problems with this approach is the lack of a 
Schr\"odinder-like equation providing an expresion of $\hat{H}$ in terms
of $x$ and ${d \over dx}$, since we have only at our disposal the 
Klein-Gordon-like equation (\ref{Klein-Gordon}) and we would need its 
``square root". In the next section we shall obtain an expression for 
$\hat{H}$ (at low order in $1/N$) in terms of $x$ and ${d \over dx}$ 
through a Schr\"odinger-like equation derived from an infinite-order 
polarization.

\section{A Perturbative Approach Involving a Perturbed Scalar Product}

Another way of approaching the $t$-factorization problem consists in
taking the ``square root'' of the second-order polarization above, a
solution to the conditions defining an (infinite-order) polarization given in a
power series $<\tilde{X}^{\infty}_{t},\tilde{X}^{L}_{p}>$, where
\begin{equation}
\tilde{X}^{\infty}_{t} = \tilde{X}^{L}_{t} + \frac{i}{\hbar} \left\{
\sqrt{\hbar^2\w^2 N(N-\lambda) - \hbar^2c^2 \left[ (\tilde{X}^{L}_{x})^2 +
m^2\w^2 (\tilde{X}^{L}_{p})^2\right] } - mc^2 \right\} \,.
\end{equation}

\ni This infinite-order polarization can be solved order by order to obtain 
a perturbative expansion for the wave functions. The first-order terms in 
$1/N$, or, equivalently, in $1/c^2$, for $\tilde{X}^{\infty}_{t}$ are 
(we are taking into account the other polarization equation, 
$\tilde{X}^{L}_{p}\Psi=0$):
\begin{equation}
\tilde{X}^{\infty}_{t} \approx \tilde{X}^{L}_{t} - 
    \frac{i\hbar}{2m}(\tilde{X}^{L}_{x})^2 - \frac{i}{4N}\left[ 
    \frac{\hbar^2}{2m^2\w} (\tilde{X}^{L}_{x})^4 - \w (1 - 2\sigma)\right]\,.
\end{equation}

\ni Here we have introduced $\lambda\equiv \frac{\sigma}{N}$ since, although 
the parameter $\lambda$ can take any value, to obtain the correct 
energy for the non-relativistic Harmonic Oscillator in the limit 
$c\rightarrow \infty$, it has to be of order lower or equal to $1/N$. 
In particular, the solutions obtained in \cite{Aldaya91}, characterized by 
$\lambda=1$, do not satisfy this requirement since the energy eigenvalues in 
the limit $c\rightarrow \infty$ are $E_n = \hbar\w n$, losing the vacuum energy 
$\medio\hbar\w$ which characterizes the quantum fluctuations of the Harmonic
Oscillator system.

Once the common $\zeta$- and $p$-dependences have
been factorized out, the new polarization gives for $\varphi(x,t)$:
\begin{eqnarray}
i\hbar\frac{\partial\varphi}{\partial t} &=&
   -\frac{\hbar^4}{8m^{3}c^{2}}\frac{\partial^{4}\varphi}{\partial
x^{4}}-
   \frac{\hbar^2}{2m}(1+\frac{3\omega^{2}x^{2}}{2c^{2}})
\frac{\partial^{2}\varphi}{\partial x^{2}}
   -\frac{\hbar^2\omega^{2}x}{2mc^{2}}\frac{\partial\varphi}{\partial
x}+\nn\\
 & &
\frac{1}{2}m\omega^{2}x^{2}(1-\frac{\omega^{2}x^{2}}{4c^{2}})
\varphi + \frac{\hbar^2\w^2}{4mc^2}(1-2\sigma)\varphi  +O(c^{-4}) 
\equiv (\hat{H}-mc^2)\varphi \, .\label{Schlikeeq}
\end{eqnarray}
\noindent (Note that we have substracted the rest mass from the Hamiltonian,
in order to get the correct non-relativistic limit). Ordinary perturbation 
theory, when applied to the Hamiltonian (\ref{Schlikeeq}), yields the correct 
energy  
$E_n\approx \hbar\omega (\frac{1}{2}+n)+\frac{\hbar\w}{8N}(1-4\sigma)
+O(c^{-4})$ and eigenfunctions
$\varphi_n(x,t)=\exp(-\frac{i}{\hbar}E_nt)\tilde{\phi}_n(x)$, where
\begin{eqnarray}
\tilde{\phi}_{n}(x)&=&\left(\frac{m\omega}{\hbar \pi}\right)^{\frac{1}{4}}
\frac{\exp(-m\omega x^2/2\hbar)}{2^{n/2}\sqrt{n!}}
\left(H_n(\xi)+\frac{1}{8N}[\frac{1}{8}H_{n+4}(\xi)+H_{n+2}(\xi)+
 \right. \nn \\
& & \left.4n(n-1)H_{n-2}(\xi)-2n(n-1)(n-2)(n-3)H_{n-4}(\xi)]
+O(c^{-4})\right) \, ,
\end{eqnarray}
\ni for any value of $\sigma$. The differences among different values of 
$\sigma$ come at order $1/c^4$, as far as $\sigma$ is of order 1. 
If $\sigma$ was of order 
$c^2$ (i.e., $\lambda$ of order 1), then the corrective terms would appear 
at order 1 in the energy and at order $1/c^2$ in the wave function, as is 
the case of 
the solutions considered in \cite{Aldaya91}, as commented before. In this 
paper we shall restrict ourselves to the case $\sigma$ of order 1 
($\lambda$ of order $1/c^2$), since it reproduces completely the 
non-relativistic limit, not only the wave functions, but also the energy.

However, $\int \d x\tilde{\phi}_n(x)\tilde{\phi}_m(x)\neq 0$ for 
$m\neq n$, the reason being that the perturbative terms added to the 
non-relativistic Hamiltonian (\ref{Schlikeeq}) are not Hermitian. 
It is important to stress that with this perturbative approach one can see
clearly why the energy wave funtions are not orthogonal to each other, 
since we have an explicit expression for the Hamiltonian (although at lower
order in powers of $1/c^2$), and this proves to be non-Hermitian.

 We could add corrective
terms to the perturbed Hamiltonian in order to make it Hermitian (note that 
the Hermiticity of $\hat{H}$ does not depend on the value of $\sigma$), but 
since the expression of $\hat{H}$ is given {\it a priori}, from a more general 
theory\footnote{The perturbed Hamiltonian has been obtained from a power series 
expansion for a higher-order polarization, but it could well had been obtained 
from a Foldy-Wouthuysen transformation of a Dirac equation for the Relativistic 
Harmonic Oscillator, as it is the case for the corrective relativistic terms
of the Hydrogen atom system \cite{Landau}.}, we are forced to associate the 
problem with the scalar product, which is no longer adequate for the 
perturbed theory. Thus, a perturbation of the scalar product itself is 
required, and this can be achieved by considering an arbitrary power
expansion in $c^{-2}$ as the perturbed measure, and determining the
corresponding coefficients by using the condition that the Hamiltonian be 
Hermitian at each 
order. As a result, we get the expression
\begin{equation}
\int\d x\{1-\omega^2x^2c^{-2}+O(c^{-4})\}
\end{equation}
\ni for the perturbed measure, restoring in this way the
unitarity of the theory. The power series in $c^{-2}$ we have obtained 
corresponds to that of $\int\d x\alpha^{-2}$, in agreement with the solution
proposed for the exact case in the previous section (see also \cite{Oscilata}
for a discussion).
 

The normalized wave functions (according to the perturbed measure) are:
\begin{eqnarray}
\Phi_n(x) &\equiv& \left(1 + \frac{2n+1}{4N}\right)
                 \tilde{\phi}_n(x) \nn \\
    &=& \phi_n(x) + \frac{1}{16N} \left[ 
 \sqrt{(n+1)(n+2)(n+3)(n+4)}\phi_{n+4}(x)   \right. \nn \\
 & & + 4\sqrt{(n+1)(n+2)}\phi_{n+2}(x) 
     +8(n+\medio)\phi_n(x) + 4\sqrt{n(n-1)}\phi_{n-2}(x) 
\label{normalized} \\
 & & \left. -\sqrt{n(n-1)(n-2)(n-3)}\phi_{n-4}(x) \right] \nn \,,
\end{eqnarray}

\ni where $\phi_n(x) = \left(\frac{m\omega}{\hbar \pi}\right)^{\frac{1}{4}}
\frac{\exp(-m\omega x^2/2\hbar)}{2^{n/2}\sqrt{n!}} H_n(\xi)$
are the normalized (non-relativistic) Harmonic Oscillator wave functions. 
It is easy to check that these wave functions coincide, up to $O(1/c^4)$, with
the exact wave functions (\ref{exactas2}) normalized with respect to the
measure $\int dx/\alpha^2$.

We can easily check that the relation between $\Phi(x)$ and $\varphi(x,t)$,
\begin{equation}
\Phi_n(x) = e^{\frac{i}{\hbar}E_nt} \left(1+\frac{n+\medio}{2N}\right)
\varphi_n(x,t)\,,
\end{equation}

\ni corresponds with the lower order terms of the transformation $\hat{U}$
given in (\ref{U}) for energy eigenfunctions. But now, when passing to 
arbitrary functions in (\ref{U2}), we have an explicit expression, at least
at lower order, for the Hamiltonian $\hat{H}$ in terms of $x$ and 
${d\over dx}$ given by (\ref{Schlikeeq}).

We think that the present perturbative approach
provides strong support to the idea that Perturbation Theory must be
in general modified in order to incorporate "non-Hermitian" 
corrections, like terms of the form $x^2\frac{d^2}{dx^2}$, which appear 
as a consecuence of the non-trivial curvature of the space-time (such as 
the Anti-de Sitter universe). Furthermore, our particular example also 
constitutes an
important step towards the solution of the general Cauchy problem in
non-globally hyperbolic spacetimes \cite{Birrell,Isham}, the 
Anti-de Sitter universe being one of the few exactly solvable cases.

\section*{Acknowledgment}

This work was partially supported by the spanish Direcci\'on General de 
Investigaci\'on Cient\'\i fica y T\'ecnica. J. Guerrero thanks the 
spanish MEC for a postdoc grant and the Dipatartimento di Scienze Fisiche, 
Universit\`a di Napoli, for its hostipality.

\end{document}